\documentclass[12pt,preprint]{aastex}

\slugcomment{Draft version \today}

\shorttitle{Primary eclipse atmospheric probing depths of exo-Earths}
\shortauthors{B\'{e}tr\'{e}mieux \& Kaltenegger}

\begin{document}


\title{Impact of atmospheric refraction: How deeply can we probe 
exo-Earth's atmospheres during primary eclipse observations?}


\author{Yan B\'{e}tr\'{e}mieux and Lisa Kaltenegger\altaffilmark{1}}
\affil{MPIA (only until September) - Looking forward to new job opportunities}
\email{betremieux@mpia.de}


\altaffiltext{1}{Harvard-Smithsonian Center for Astrophysics, 
60 Garden street, Cambridge MA 02138, USA}


\begin{abstract}

Most models used to predict or fit exoplanet transmission spectra do not include 
all the effects of atmospheric refraction. Namely, the angular size of the star with respect to the
planet can limit the lowest altitude, or highest density and pressure, probed during primary 
eclipses, as no rays passing below this critical altitude can reach the observer.
We discuss this geometrical effect of refraction for all exoplanets, and tabulate the critical altitude, 
density and pressure for an exoplanet identical to Earth with a 1~bar N$_2$/O$_2$ atmosphere, as a function of 
both the incident stellar flux (Venus, Earth, and Mars-like) at the top of 
the atmosphere, and the spectral type (O5-M9) of the host star. We show that such a habitable exo-Earth
can be probed to a surface pressure of 1~bar only around the coolest stars. We present 0.4-5.0~$\mu$m 
model transmission spectra of Earth's atmosphere viewed as a transiting 
exoplanet, and show how atmospheric refraction modifies the transmission spectrum depending on 
the spectral type of the host star. We demonstrate that refraction is another phenomenon 
that can potentially explain flat transmission spectra over some spectral regions.

\end{abstract}


\keywords{atmospheric effects --- Earth --- line: identification --- methods: analytical --- 
planets and satellites: atmospheres --- radiative transfer}



\section{Introduction}

The on-going discovery of new exoplanets and its potential for finding other Earth-like planets, has fueled 
studies (Ehrenreich et al.~2006; Kaltenegger \& Traub~2009; Pall\'{e} et al.~2009; 
Vidal-Madjar et al.~2010; Rauer et al.~2011; Garc\'ia Mu\~{n}oz et al.~2012; Snellen et al.~2013; Hedelt et al.~2013; 
B\'etr\'emieux \& Kaltenegger~2013; Rodler \& L\'opez-Morales~2014; Misra et al. 2014) modeling the spectrum 
of the Earth's atmosphere viewed as a transiting exoplanet. However, most models predicting, or fitting, the spectral 
dependence of planetary transits have not fully included a fundamental phenomenon of the atmosphere on radiation: refraction.
As light rays traverse an atmosphere, they are bent by refraction from the major gaseous 
species. The angular deflection is, to first order, proportional to the density of the gas (Goldsmith 1963), 
so that the greatest ray bending occurs in the deepest atmospheric regions. This differential refractive bending 
of light with altitude generally decreases the flux of a point source when observed through an atmosphere. 
Combined with absorption and scattering from gas and aerosols, it has been used to interpret stellar 
occultation by planetary atmospheres in our solar system to determine their composition with altitude
(see, e.~g., Smith \& Hunten~1990, and references therein). 

In an exoplanetary transit geometry, where the star is an extended source with respect to the observed 
planetary atmosphere and the observer is infinitely far away, refraction produces different effects. 
Right before and after a transit, some stellar radiation can bend around through the deeper regions of the 
planet's opposite limb, toward the observer, while the stellar disk is unocculted. This results in a global 
flux increase if the atmosphere is not opaque (Sidis \& Sari~2010). Conversely, 
during transit, refraction from the deeper atmospheric regions deflects light away from the observer.
The latter effect can limit how deeply atmospheres can be probed during exoplanetary transits. Indeed, when the 
angular extent of the star with respect to the planet is sufficiently small, rays traversing a planetary atmosphere 
below a critical altitude will be bent so much that no rays from the stellar surface can reach the distant observer. 
This effect masks molecular absorption features that originate below this altitude. 
Whereas refractive effects have been shown not to be important for hot giant exoplanets (see
the analysis for HD 209458b by Hubbard et al. 2001), and the high-altitude hazes and clouds in the Venusian 
atmosphere are located above the critical altitude of a Venus-Sun analog (Garc\'ia Mu\~noz \& Mills 2012),
for an Earth-Sun analog, the critical altitude occurs in the upper troposphere, above the cloud deck, and 
the strength of water absorption features are substantially 
reduced (Garc\'ia Mu\~noz et al.~2012, B\'{e}tr\'{e}mieux \& Kaltenegger~2013).

How deeply can we probe the atmosphere of Earth-like planets in transit around other stars? For which stellar spectral 
types can one probe down to the planetary surface? What impact does it have on the spectral 
features of an exo-Earth during primary eclipse observations? In this paper, we address these questions
for an exo-Earth, a planet defined here as having the same mass and radius, and an 
identical 1 bar N$_2$/O$_2$ atmosphere, as the Earth. We
explore how the critical altitude changes with spectral type, along the Main sequence,
of the host star (O5-M9), modifying the planet-star distance to keep the incident stellar flux constant, for 
fluxes between that of present-day Venus and Mars. We first discuss the basic effects of refraction in an exoplanet 
transit geometry (Section~\ref{theory}), present our refraction model and our results (critical altitudes, densities, 
and pressures) around various stars (Section~\ref{model}), and then show its impact on an exo-Earth's
transmission spectrum from 0.4 to 5.0~$\mu$m (Section~\ref{spectrum}), a spectral region relevant to the 
James Webb Space Telescope (JWST), as well as ground-based observatories like E-ELT.

\section{Refraction theory for exoplanet transit geometries} \label{theory}

\subsection{Atmospheric refraction} \label{atmrefsect}
\begin{figure}
\epsscale{1.0}
\plotone{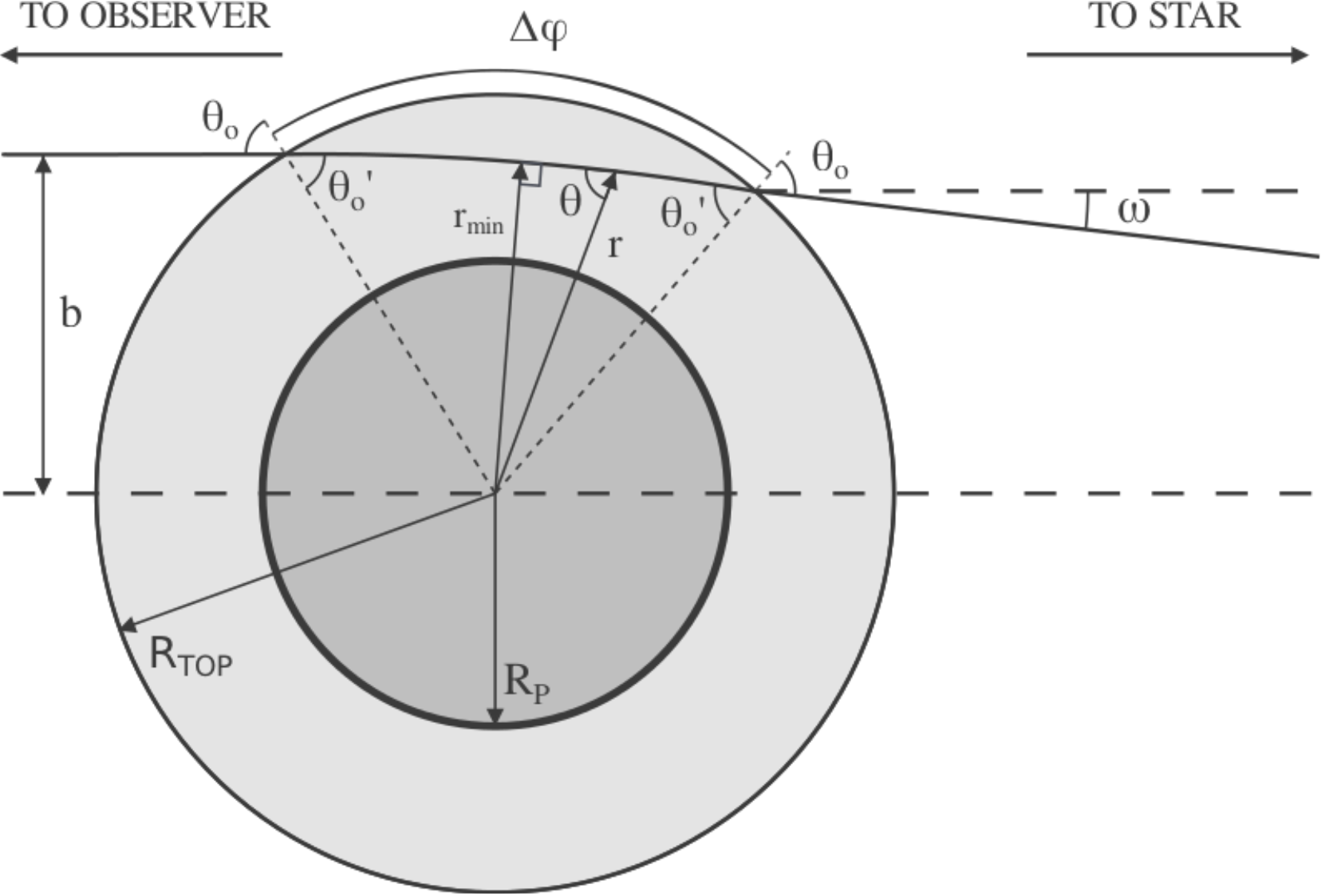}
\caption{Bending of light ray by atmospheric refraction for 
an exoplanetary transit. The planetary body (dark grey) and the atmosphere (light grey) are shown, 
along with the observer-to-planetary-center axis (dashed line) with the observer to the left and the star 
to the right. The radius of the planetary surface, R$_p$, the top of the atmosphere, R$_{top}$, 
and the zenith angle, $\theta$, of the ray for a given radial position, r, are 
also indicated. A light ray observed with an impact parameter~b, reached a 
minimum radial position r$_{min}$, and is deflected by $\omega$ by the atmosphere. 
\label{fig1}}
\end{figure}

The exponentially increasing density, and hence refractivity, of the atmosphere with depth causes 
light rays to bend toward the surface as they traverse the atmosphere (see Figure~\ref{fig1}). 
Rays follow a path described 
by an invariant $L = (1 + \nu(r)) r \sin\theta(r)$  where both the zenith angle of a ray, $\theta(r)$, 
and the refractivity of the atmosphere, $\nu(r)$, are functions of the 
radial position of the ray with respect to the center of the planet. The refractivity of the atmosphere 
depends on its composition, and is given by
\begin{equation} \label{refrac}
\nu(r) = \left( \frac{n(r)}{n_{STP}} \right) \sum_{j} f_{j}(r) {\nu_{STP}}_{j} = \left( \frac{n(r)}{n_{STP}} \right) \nu_{STP}(r) ,
\end{equation}
where $n(r)$ is the number density, $n_{STP}$ is the number density at standard temperature and pressure
(STP) also known as Loschmidt's number, $f_{j}(r)$ is the mole fraction of 
the j$^{th}$ species, ${\nu_{STP}}_{j}$ is the STP refractivity of the j$^{th}$ species, while $\nu_{STP}(r)$ 
is the STP refractivity of the atmosphere. Note that the ratio inside the parenthesis in Equ.~\ref{refrac} 
is the number density expressed in units of amagat, which we will use throughout the paper. 
Assuming the refractivity just outside the atmosphere is 
zero, we can relate the lowest altitude reached by a ray, $\Delta z_{min}$, to its impact parameter, $b$. 
From the conservation of the invariant $L$, as well as the geometry in Figure~\ref{fig1} combined with Snell's law,
\begin{mathletters} 
\begin{eqnarray}
r_{min} = \Delta z_{min} + R_{p} \label{refpath2} \\ 
L = (1 + \nu(r_{min})) r_{min} = (1 + \nu(R_{top})) R_{top} \sin\theta'_{0} \label{refpath1} \\ 
b = R_{top} \sin\theta_{0}  = (1 + \nu(R_{top})) R_{top} \sin\theta'_{0} \label{refpath3}
\end{eqnarray}
\end{mathletters}
we can see that the impact parameter is equal to the ray invariant $L$. 
Here, $R_{top}$ is the radius of the top of the atmosphere, $R_{p}$ is the radius of the planetary 
surface, $r_{min}$ is the lowest radial position reached by a ray, while $\theta_{0}$ and $\theta'_{0}$ are 
the zenith angle of the ray just outside and inside the top atmospheric boundary, respectively (see Figure~\ref{fig1}). 

The deflection of the ray, $\omega$, expressed in radian, is given by 
\begin{equation} \label{deflection}
\omega = \Delta \phi + 2 \theta_{0} - \pi
\end{equation}
where $\Delta \phi$ can be described as an angular travel of the ray through the atmosphere. For a simple
homogeneous isothermal atmosphere, Goldsmith (1963) showed that to first order the deflection of the ray 
is given by
\begin{equation} \label{defexp}
\omega =  \left( \frac{{2 \pi r_{min}}}{H(\Delta z_{min})} \right)^{1/2} \nu_{STP}(\Delta z_{min})  n(\Delta z_{min})
\end{equation}
where $H$ is the atmospheric density scale height. Since the density decreases exponentially with 
altitude, the ray deflection is very sensitive to the lowest altitude reached and thus to the impact 
parameter of the ray. Note that since most atmospheres are far from isothermal, we use a ray-tracing 
algorithm (see Section~\ref{modelsect}, and B\'etr\'emieux \& Kaltenegger 2013) to compute the atmospheric 
deflection.

\subsection{Critical deflection of a planet-star system} \label{critdefsect}

The exoplanet transit geometry is exactly the reverse of the stellar occulation geometry. In stellar 
occultations, the source is infinitely far away from the refractive medium, or planetary atmosphere, 
while the observer is relatively close. In exoplanet transits, the observer is infinitely far away 
while the source is relatively close to the atmosphere. In this geometry, where all the rays reaching 
the observer are parallel with varying impact parameters, refraction produces three effects. 
The most obvious one, also present in stellar occultations,
is an increase in the optical depth, due to an increase in the ray's path length through the 
atmosphere as the ray is bent. Previous calculations usually incorporate this effect when refraction 
is said to be included (see, e.g., Kaltenegger \& Traub 2009, Benneke \& Seager 2012).
The second effect comes from the mapping of $r_{min}$ 
into $b$. As we can see from Equations~\ref{refpath1} and~\ref{refpath3}, $b$ is always larger than $r_{min}$. 
The difference between the two quantities increases with refractivity, or decreasing altitude. 
The largest difference exists for the rays grazing the planetary surface ($r_{min} = R_p$), 
thus the planet appears slightly larger to a distant observer. This second effect should be implicitly 
included with the first one when optical depth calculations are done as a function of $b$, 
rather than $r_{min}$.  

\begin{figure}
\epsscale{0.7}
\plotone{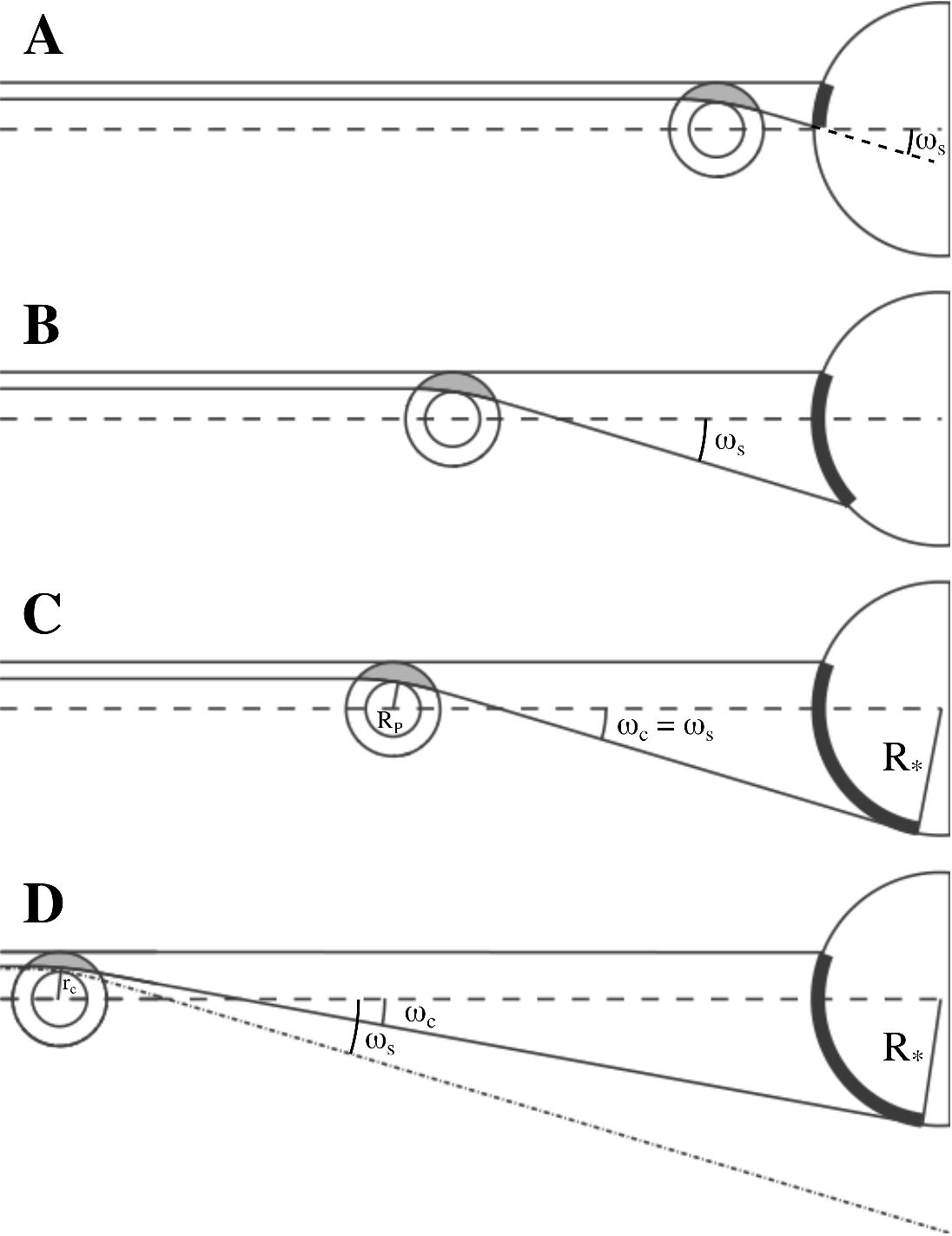}
\caption{Geometry of exoplanetary transits with increasing 
planet-star distance. Two solid lines, joining the star (right) to the 
observer (left along the dashed line), define the envelope of rays that can be 
seen by the observer through one exoplanetary limb. Shaded areas show the atmospheric region 
that can be probed by the observer, while the thick circumference shows the stellar region where 
the rays come from. The stellar radius, R$_{\ast}$, radius of planetary surface, R$_p$, critical 
radius, r$_c$, surface deflection, $\omega_s$, critical deflection, $\omega_c$ are indicated.
With short planet-star distances, light rays come from a stellar region
behind the planet (panel A). As the planet-star distance increases, some rays come from region of the star beyond
the planet's opposite limb (panel B). At a critical planet-star distance, rays grazing the planetary surface 
come from the star's opposite limb (panel C). For larger planet-star distances, only rays that traverse the 
atmosphere above a critical altitude can reach the observer, and lower atmospheric regions
cannot be probed (panel D). 
\label{fig2}}
\end{figure}

The third effect, illustrated in Figure~\ref{fig2} for the cylindrically symmetric case when the 
planet occults the central region of its star, is a purely geometrical effect
linked to the angular size of the host star with respect to the planet, and has only 
been included in a few papers (Garc\'ia Mu\~noz \& Mills~2012; Garc\'ia Mu\~noz et al.~2012; 
B\'{e}tr\'{e}mieux \& Kaltenegger~2013; Rodler \& L\'opez-Morales~2014; Misra et al. 2014).
At the top of the atmosphere, the ray from the star is undeflected. 
As the impact parameter of the rays decreases, the deflection of the rays increases. A maximum 
deflection occurs for rays that graze the planetary surface. Although the deflection is small, 
the planet-star distance is large, so that the lateral displacement of the rays is significant. 
If the planet is sufficiently close to its star, or the atmosphere is sufficiently tenuous, 
rays reaching the observer after passing through the planetary atmosphere, merely come from a wider annular 
region on the stellar surface than the simple projection of the planetary atmosphere on the star 
would suggest (see panel~A in Figure~\ref{fig2}). As the distance between the planet and the star 
increases, the inner edge of this region moves toward and eventually crosses the center of the star, and 
continues further on radially outward. Eventually, this edge can move beyond the projection of the opposite 
limb of the planet on the stellar surface (panel~B). At a critical planet-star distance, rays grazing the 
planetary surface reach the opposite limb of the star (panel~C). If the planet orbits beyond this 
critical distance, then only rays above a critical altitude can traverse the atmosphere and reach the 
observer (panel~D). Atmospheric regions below that altitude cannot be probed, 
therefore the planet will seem larger than it is. Atmospheric opacity does 
limit the depth that can be probed as well, but for the sake of brevity, we only refer to the 
effect from refraction throughout this paper unless specified otherwise. Note that in Section~\ref{spectrum},
we present all effects, not only refraction, and therefore produce 
the transmission spectrum of Earth seen as an exoplanet in transit around different host stars.

Whether the lower atmospheric regions are dark due to
this refraction effect or because these regions are intrinsically opaque makes no difference to the transmission 
spectrum. Refraction does not change the transmission spectrum of an Earth-Sun analog shortward of 
0.4~$\mu$m because of the high opacity of the atmosphere in the ultraviolet (B\'etr\'emieux \& Kaltenegger~2013).
Figure~\ref{fig2} shows that stellar limb darkening effects will probably be different from what 
has previously been modeled, because the stellar region of origin of the rays is not merely the 
projected limb of the atmosphere on the star. This could explain part of the discrepancy
between theoretically-derived limb darkening coefficients and those inferred through light-curve 
fitting (see discussion in Csizmadia et al. 2012 for other explanations). However, to focus on the effects of refraction, 
we ignore limb darkening in this paper and assume that the specific intensity across the stellar disk is uniform.

One can calculate the critical deflection, $\omega_c$, 
undergone by a ray that reaches the opposite stellar limb if one assumes that all of the bending of the 
ray occurs at the minimum altitude reached by the ray. Given that the path length of the ray through 
the atmosphere is much smaller than the planet-star distance, $d$, and that the curvature of that path is 
very small, this approximation introduces negligible errors when computing 
the total lateral displacement of the beam at the star's position. In that case, from the 
geometry in panel D (Figure~\ref{fig2}), we can see that
\begin{equation} \label{critdef}
\sin\omega_{c} = \frac{r_c + R_{\ast}}{d} = \frac{(R_p + \Delta z_c) + R_{\ast}}{d} 
\approx \frac{R_p + R_{\ast}}{d}
\end{equation}
assuming that the critical altitude, $\Delta z_c$, is much smaller than the sum of the planetary surface 
radius, $R_p$, and the stellar radius, $R_{\ast}$. 
The critical altitude is then defined as the altitude at which the atmospheric ray deflection, dependent on the atmospheric 
properties and size of the exoplanet (Equ.~\ref{defexp}), matches the critical deflection of the planet-star system.
The critical density and critical pressure are the number density and pressure of the atmosphere at the 
critical altitude, respectively. Note that when $R_p \ll R_{\ast}$, Equation~\ref{critdef} reduces to the expression 
for the angular radius of the star with respect to the planet. The critical deflection is mostly sensitive to the $R_{\ast}/d$ ratio, 
and does not strongly depend on the atmospheric properties of the exoplanet.
A change in planetary radius from 1$R_{\Earth}$ (Earth) to 2$R_{\Earth}$ (super-Earth) changes the critical deflection by 
about 1\% around a Sun-like star, and about 10\% around a M9 star. 

The assumed geometry is only strictly valid when the observer, planet and star are perfectly 
aligned. Deviation from this alignment, during the course of a transit, breaks the 
symmetry of the two opposing limbs, where radiation going through one limb can penetrate deeper 
than in the opposite one. Just outside of transit (OT), the planetary limb toward the star cannot be probed 
at all, while the critical deflection for the opposite limb, $\omega_{cOT}$, increases to
\begin{equation} \label{critdefOT}
\sin\omega_{cOT} = \frac{r_c + R_{top} + 2R_{\ast}}{d} 
\approx \frac{2R_p + 2R_{\ast}}{d} = 2\sin\omega_{c} 
\end{equation}
Only a detailed study of the refraction effect on the full exoplanetary transit light-curve will reveal to 
what extent this modifies the wings of the transit light-curve. For this paper, we will use this simpler
geometry as an average representation to discuss the first-order effects of refraction on the 
exoplanetary transmission spectrum.

\subsection{Dependence of critical deflection with stellar spectral type}

The wavelength-integrated stellar flux, $F_{\ast}$, on the top of a planetary atmosphere is given, 
in a first-order approximation assuming the star to be a blackbody, by 
\begin{equation} \label{flux}
F_\ast = \sigma T_{\ast}^4 \left( \frac{R_{\ast}}{d} \right)^2
\end{equation}
where $T_{\ast}$ is the stellar effective temperature, 
and $\sigma$ is the Stefan-Boltzmann constant. Hence, the planet-star distance, $d$, at which the stellar 
flux on top of an exoplanet's atmosphere equals that of a solar system planet at a distance $d_{\Sun}$, is given by
\begin{equation} \label{distanceratio}
\frac{d}{d_{\Sun}} =   \left( \frac{R_{\ast}}{R_{\Sun}} \right) \left( \frac{T_{\ast}}{T_{\Sun}} \right)^2
\end{equation}
where here, the subscript $\Sun$ refers to quantities for our solar system. 

Using Equations~\ref{critdef} and~\ref{distanceratio}, 
the critical deflection around other stars can be related to that of planets in our solar system, $\omega_{c\Sun}$, by 
\begin{equation} \label{deflratio}
\frac{\sin \omega_c}{\sin \omega_{c\Sun}} =   \left( \frac{R_p + R_{\ast}}{R_p + R_{\Sun}} \right) \left( \frac{d}{d_{\Sun}} \right)^{-1}
 =  \left[ \left( \frac{R_p + R_{\ast}}{R_p + R_{\Sun}} \right)  \left( \frac{R_{\Sun}}{R_{\ast}} \right) \right] 
\left( \frac{T_{\Sun}}{T_{\ast}} \right)^2 .
\end{equation}
The ratio inside the square bracket simplifies to unity when $R_p \ll R_{\ast}$. Even in the case of an Earth-sized planet 
around an M9 star, the ratio is still 1.10, fairly close to unity. Thus, in the small angle approximation and 
for $R_p \ll R_{\ast}$, Equation~\ref{deflratio} simplifies to
\begin{equation} \label{deflratiosimp}
\frac{\omega_c}{\omega_{c\Sun}} \simeq  \left( \frac{T_{\Sun}}{T_{\ast}} \right)^2 
\end{equation}
and is independent of stellar radius. Hence, for exoplanets with the same incident stellar flux, the critical 
deflection of the planet-star system is larger around cooler stars, and deeper regions of the planetary atmosphere 
can be probed.

Note that Equ.~\ref{deflratiosimp} is not used to determine the critical deflections of the 
planet-star system, from which the critical altitudes presented in Section~\ref{results} are derived. Rather, we 
first use Equ.~\ref{distanceratio} to compute the planet-star distance, and then Equ.~\ref{critdef} to determine
the corresponding critical deflection. All of the atmosphere can be probed when the critical deflection of the 
planet-star system exceeds the ray deflection at the surface of the planet.

\section{Critical altitudes and densities} \label{model}

\subsection{Model description} \label{modelsect}

\begin{figure}
\epsscale{1.0}
\plotone{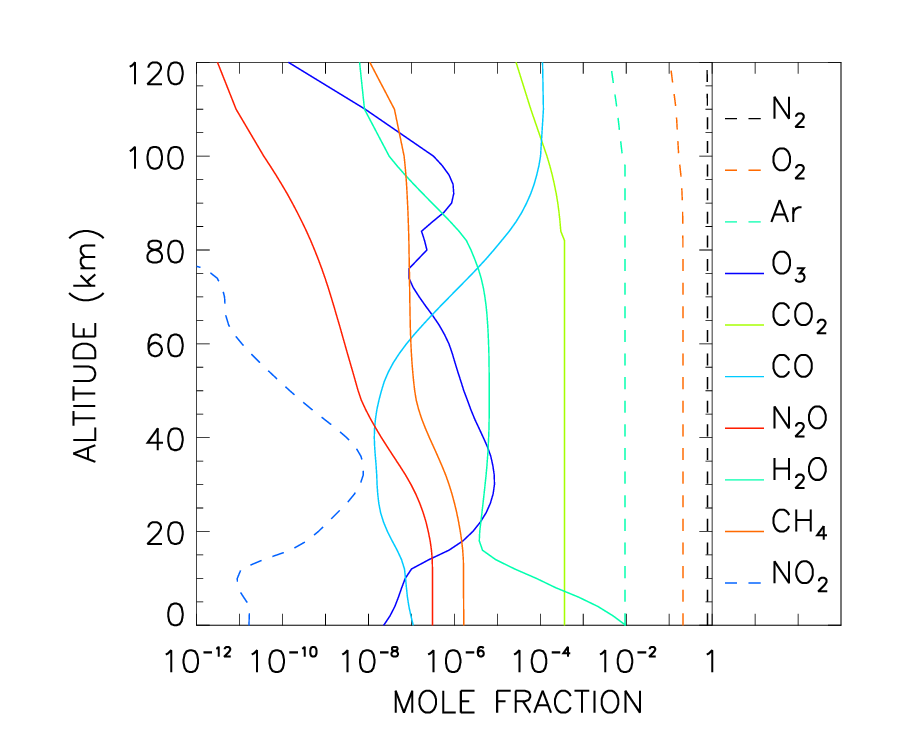}
\caption{Composition of the Earth's atmosphere as a function of altitude for solar minimum conditions
(see B\'etr\'emieux \& Kaltenegger 2013).
\label{fig3}}
\end{figure}

To compute the refractive and absorptive properties of an exo-Earth, we use the disc-averaged Earth solar 
minimum model atmosphere and the ray tracing and radiative transfer model described in 
B\'etr\'emieux \& Kaltenegger (2013). This model traces rays through a spherical
atmosphere discretized in 0.1~km thick homogeneous layers and computes $\Delta \phi$ from which 
the deflection is derived with Equation~\ref{deflection}. This is done as a function of the 
minimum altitude reached by the ray, $\Delta z_{min}$, from which $b$ is computed with Equations~\ref{refpath2}, 
\ref{refpath1} and~\ref{refpath3}. The model also computes column abundances and average 
mole fractions for different species along the path of the ray, from which the optical depth and the transmission 
are derived. In this paper, the top of the atmosphere, $R_{top}$, is chosen 120~km above $R_p$ where atmospheric absorption 
and refraction are negligible across our spectral region of interest (0.4~-~5~$\mu$m). The composition of 
Earth's atmosphere with altitude is shown in Figure~\ref{fig3}. The main refractors in the Earth's atmosphere are N$_2$, O$_2$, Ar, 
and CO$_2$, which make-up more than 99.999\% of the atmosphere. Although we include all these in our model, ignoring 
Ar and CO$_2$ would create less than a 1\% change in the refractivity, and hence on ray deflection. 
The contribution from water vapor is even smaller and is ignored here. Since the main refractors are well mixed 
throughout the atmospheric region considered, the STP refractivity hardly changes with altitude, and 
is assumed to be the planet's surface value, $\nu_{STP}$, at all altitudes throughout our calculations.

\begin{figure}
\epsscale{1.0}
\plotone{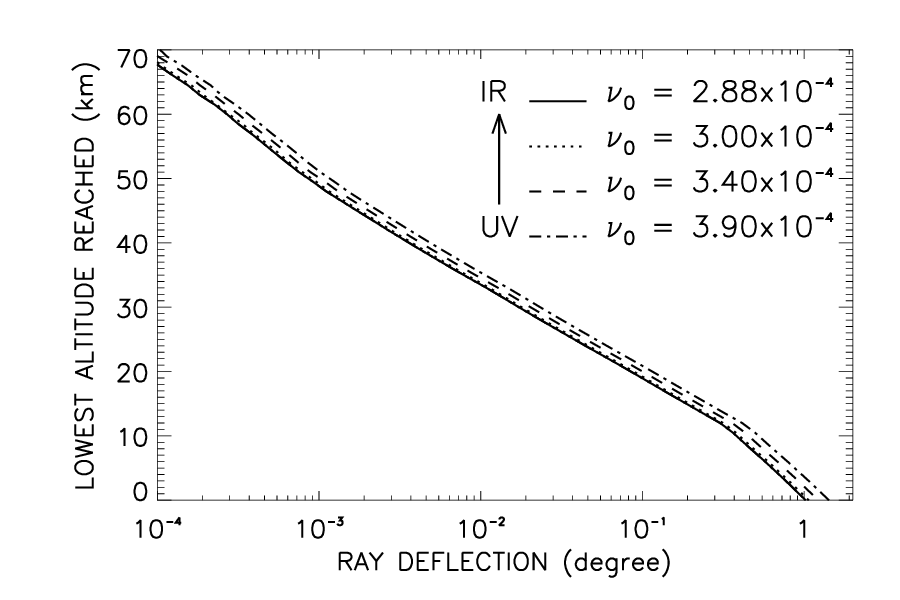}
\caption{Ray deflection caused by atmospheric refraction as a function of the lowest altitude reached by that 
ray. The various curves show results for a sample of STP refractivities of Earth's atmosphere,~$\nu_{STP}$, 
from the ultraviolet to the infrared, assumed to be constant with altitude (see text). 
\label{fig4}}
\end{figure}

As the STP refractivity varies with wavelength, the ray deflection and critical altitude will also be 
wavelength dependent. However, $\nu_{STP}$ changes only slightly in the infrared~(IR), going from 2.8798$\times10^{-4}$ 
at 30~$\mu$m to 2.88$\times10^{-4}$ at 8.8~$\mu$m. Its variation with wavelength, although greater, is still
small from the infrared to the near-infrared (NIR) and the visible~(VIS). $\nu_{STP}$ is 2.90$\times10^{-4}$ 
at 0.87~$\mu$m, and 3.00$\times10^{-4}$ at 0.37~$\mu$m. In the ultraviolet (UV), $\nu_{STP}$ changes 
more rapidly, reaching 3.40$\times10^{-4}$ at 0.20~$\mu$m, and becomes larger than 3.90$\times10^{-4}$ for 
wavelengths shorter than 0.14~$\mu$m. Figure~\ref{fig4} shows the ray deflection as a function of the lowest altitude 
reached by a ray, for different $\nu_{STP}$ spanning values covering the IR to the UV. The difference in the 
STP refractivity from the IR to the UV changes the lowest altitude reached by rays, which have undergone 
the same deflection, by about 2~km for atmospheric regions above 12~km, and about 3.5~km 
closer to the surface. The difference in the STP refractivity across our spectral region of 
interest (IR-VIS), induces a change in a ray's minimum altitude of only about 0.25~km, most of which occurs 
across the visible. Since that difference is small, we will use a single value of 2.88$\times10^{-4}$ for $\nu_{STP}$ 
throughout the rest of the paper. Note that for this refractivity value, using Equ.~\ref{defexp} instead of 
our ray tracing model underestimates the ray deflection by up to approximately 4\% near the surface, while 
it overestimates it by an average of 7\% above 10~km altitude. Even though, in this particular case, the difference 
between the simplified isothermal treatment and our ray tracing model is small, on physical grounds the ray tracing 
model is still preferable.

\subsection{Results} \label{results}

\begin{figure}
\epsscale{1.0}
\plotone{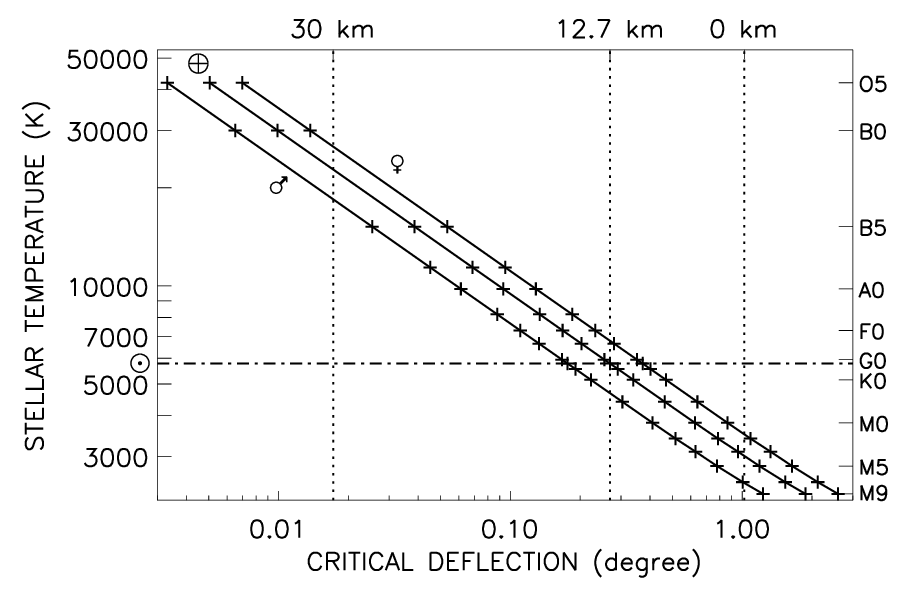}
\caption{Critical deflection, $\omega_c$, as a function of effective stellar temperature, T$_\ast$, for an Earth-sized exoplanet 
orbiting a Main sequence star (O5-M9). The three tracks, from left to right, show results for an exoplanet that is at a distance 
from its star where it receives the same stellar flux as present-day Mars, Earth, and Venus, respectively. 
Crosses show data points listed on Table~\ref{tbl_star}, with
our Sun's temperature highlighted by the dot-dashed line. Vertical dotted lines show the deflection at Earth's surface, 
at 12.7, and 30~km altitude.
Only for critical deflections greater or equal to that of the 0~km line can observers probe 
down to the exo-Earth's surface.
\label{fig5}}
\end{figure}

To determine around which spectral type stars, the atmosphere of a habitable exo-Earth can be probed down to the 
surface, we must first determine the critical deflection of the exo-Earth-star system. We compute the critical deflection of an 
Earth-sized planet ($R_p =$~6371~km) orbiting various stars along the Main sequence track, using Equations~\ref{critdef} 
and \ref{distanceratio}. The radius and effective temperature of the stars considered are listed in Table~\ref{tbl_star}. 
We do the calculations for planets which receive the same stellar flux as Earth ($d_{\Sun} = $~1~AU), and contrast 
them with planets which receive the same stellar flux as Venus ($d_{\Sun} = $~0.723~AU) and Mars ($d_{\Sun} = $~1.524~AU). 
The results are shown in Figure~\ref{fig5} along with the ray deflection for three benchmark altitudes on Earth: that of the 
surface (0~km), the Earth-Sun critical altitude (12.7~km, see B\'etr\'emieux \& Kaltenegger 2013), and at 30~km, close to the 
peak of the ozone vertical profile. Transmission spectroscopy can probe the atmosphere of an exo-Earth down to its surface for 
stars that are cooler than or the same temperature as the limit defined by the surface ray deflection. 
This limit occurs for M1, M4, and M7 stars for Venus-like, Earth-like, and Mars-like fluxes, respectively.
Hence, for Main sequence stars, observers can probe an exo-Earth to a surface pressure of 1 bar only around M-dwarfs. 
For thinner N$_2$/O$_2$ atmospheres, this limit shifts to hotter stars, while for thicker N$_2$/O$_2$ atmospheres, 
it shifts to cooler stars.

For each exo-Earth-star system with a specified incident stellar flux (Venus, Earth, or Mars-like), we determine the 
critical altitude, density, and pressure by first computing the atmospheric 
deflection as a function of a ray's minimum altitude on a 1.5~km altitude grid, and determine which region 
produces deflections closest to the system's critical deflection. We redo this step with successively finer altitude grids
across the atmospheric region which has deflections closest to the critical deflection until we get to a 0.001~km grid. 
We then determine which altitude on that grid produces a deflection closest to the system's critical deflection. 
The critical density and critical pressure are the number density and pressure of the atmosphere at this critical 
altitude, respectively.

As can be seen in Table~\ref{tbl_critalt}, whereas an exo-Earth's surface conditions can be probed around some 
of the cooler M-dwarfs, the critical altitude increases as one observes planets around hotter stars, 
and the corresponding critical density and pressure decrease.  
For an Earth-Sun analog, most of the troposphere is inaccessible to observers. Although tropospheric water content 
cannot be determined, this also means that tropospheric clouds have no impact on the transmission spectrum, and hence
they cannot be responsible for transmission light-curve variability. For stars hotter than B0 stars, the critical altitude 
lies even above 30~km, and only the mesosphere and 
thermosphere can be probed. For a given star, the critical altitude decreases as the stellar flux received 
by the exo-Earth increases because the star's angular size with respect to the planet is larger.
Although it is unclear at this point whether
a planet in the habitable zone of O and B stars could, because of the extreme UV environment or of the short lifetime
of these stars, ever develop a stable atmosphere, we have nevertheless included the calculations for exo-Earth's orbiting 
these stars to show what the extreme values are.

\subsection{Effects of different atmospheric profiles} \label{sedsect}

Although differences in the spectral energy distribution of stars lead to differences
in the thermal and chemical composition profile of an exoplanet's atmosphere (see, e.g., Rugheimer et al. 2013), 
we use an Earth profile here to focus only on the geometrical effect of refraction. Hence, the values in 
Table~\ref{tbl_critalt} illustrate only the impact of refraction. Increasing the proportion of 
UV flux would increase the photodestructive rate of N$_2$ and O$_2$ in the thermosphere, where densities are low and 
refraction is not important. In the lower atmosphere, we expect no changes in the abundances of N$_2$ and O$_2$
because convection would replenish any altitude-localized depletion. Thus, $\nu_{STP}$ would not change significantly 
in the lower atmosphere. However, increasing UV fluxes also increases the amount of ozone produced in the stratosphere 
with a resulting increase in stratospheric temperatures. Increasing visible and infrared flux increases the temperature 
in the troposphere. Hence, as the peak of the stellar spectral energy distribution shifts toward shorter wavelengths 
(i.e. hotter stars), stratospheric temperatures increase while tropospheric temperatures decrease. Since the ray deflection 
is sensitive mostly to the density of the lowest altitude reached by the ray, when the scale height doubles, the altitude 
which produces a given ray deflection will also double. Hence, to first order, the critical altitude scales with scale height. 
To second-order, since $\omega \propto H^{-1/2} \approx T^{-1/2}$ (for isothermal region of atmosphere), as T increases, 
$\omega$ decreases and the critical altitude decreases slightly.
Therefore, as T increases, the critical altitude will be slightly smaller than a simple scaling with scale height suggests. 
As the stellar spectral type changes from K to F, Rugheimer et al. (2013) showed that below 25~km altitudes, temperatures do 
not change by more than 10\% (and typically far less than that, see their Figure~3), and so neither will the critical altitudes.

\subsection{Effects of different atmospheric composition and planetary sizes}

Since atmospheric ray deflection depends on the composition of the atmosphere, as well as the radius of the planet 
(see Equation~\ref{defexp}), CO$_2$ and H$_2$/He dominated atmosphere are affected differently by the geometrical effect 
of refraction. Although the critical altitude is a useful quantity when discussing the transmission spectrum 
(see Section~\ref{spectrum}), it is linked to the critical density, $n_c$, the density which bends rays by $\omega_c$.
To first order, for an isothermal atmosphere, we can see from Equation~\ref{defexp} that
\begin{equation} \label{critdens}
n_c =  \frac{\omega_c}{\nu_{STP}(\Delta z_c)} \left( \frac{H(\Delta z_c)}{{2 \pi r_{min}}} \right)^{1/2}
= \frac{\omega_c}{\nu_{STP}(\Delta z_c)} \left( \frac{k_B T}{2 \pi m g r_{min}} \right)^{1/2}
\end{equation}
where $k_B$ is the Boltzmann constant, and $T$, $m$ and $g$ are the temperature, average molecular mass, and
gravitational acceleration of the atmosphere at $r_{min}$. Note that since the critical altitude is usually 
much smaller than the planetary radius, $r_{min}$ can be approximated by $R_p$ for the purpose of the following discussion.
For gaseous planets, the surface planetary radius is defined at a specific pressure. Compare the critical densities of 
planets with identical radii with a pure CO$_2$ (close to Venus-like), an Earth-like, and a Jupiter-like 
atmosphere. Around 0.65~$\mu$m, the refractivities of H$_2$ and He are about 1.39$\times10^{-4}$ and 3.48$\times10^{-5}$ 
(Leonard 1974). Jupiter's atmosphere, composed of 86.2\% H$_2$ and 13.6\% He (Lodders \& Fegley 1998), has a STP
refractivity of 1.25$\times10^{-4}$, significantly lower than the Earth's N$_2$/O$_2$ atmosphere (2.92$\times10^{-4}$), 
itself much lower than a pure CO$_2$ (4.48$\times10^{-4}$) atmosphere. This difference in bending power is increased 
further because a H$_2$/He mixture has much lower average mass than an N$_2$/O$_2$ mixture, which is lighter than CO$_2$. 
Hence, CO$_2$ atmospheres bend light rays the most, and for a given critical deflection of a planet-star system, 
observers can only probe to critical densities 0.53 times that of an Earth-like atmosphere. 
H$_2$/He Jupiter-like atmospheres bend light rays the least, and observers can probe denser regions with critical 
densities 8.3 times that of an Earth-like atmosphere. Therefore, mini-Neptunes with a H$_2$/He mixture can be probed 
to higher densities than Super-Earths with a N$_2$/O$_2$ or CO$_2$ atmosphere, assuming similar planetary radii. 

The size of a planet also has an impact on the critical density. Ignoring gravity, Equation~\ref{critdens} shows that 
a super-Earth twice the size as Earth can only be probed to a critical density 0.71 times that of the Earth. If one assumes that the 
solid body of the planet has a constant mass density, then the gravitational acceleration scales with the radius of the planet, 
and $n_c \propto 1/R_p$. In this case, doubling the radius of the planet halves the critical density. Therefore, super-Earth
atmospheres cannot be probed to regions as dense as Earth-sized planets.
The large size of gaseous planets relative to terrestrial ones can compensate for the 
low bending power of H$_2$/He atmospheres. Factoring in the gravitational acceleration (24.8~m/s$^2$ for Jupiter and 9.82~m/s$^2$
for Earth), and the planet's radius (71492~km for Jupiter and 6371~km for Earth), a Jupiter-sized planet with similar temperatures
as the Earth can be probed down to a critical density only 1.6 times that of the Earth.

\section{Impact of refraction on transmission spectrum} \label{spectrum}

To illustrate the impact that refraction has on the transmission spectrum of an exoplanet, we compute the 
transmission spectrum for exo-Earths orbiting different stars. We do this for exo-Earths receiving 
the same incident stellar flux as the Earth and divide the atmosphere into 80 layers
evenly spaced in altitude from the top of the atmosphere (120~km altitude) to the critical altitude 
corresponding to each star (see Table~\ref{tbl_critalt}). The bottom of each layer defines the minimum 
altitude reached by a ray, for which we can compute the refractivity, the impact parameter, and 
the transmission through the atmosphere. We compute the transmission on a 0.05~cm$^{-1}$ grid from 
2000 to 25000~cm$^{-1}$ (0.4-5.0~$\mu$m), which is then rebinned every 4.0~cm$^{-1}$ for display. 
We characterize the transmission spectrum of an exo-Earth by its effective atmospheric thickness
using the method described in B\'etr\'emieux \& Kaltenegger (2013), which combines the effects of
atmospheric opacities and refraction.
 
\begin{figure}
\epsscale{1.0}
\plotone{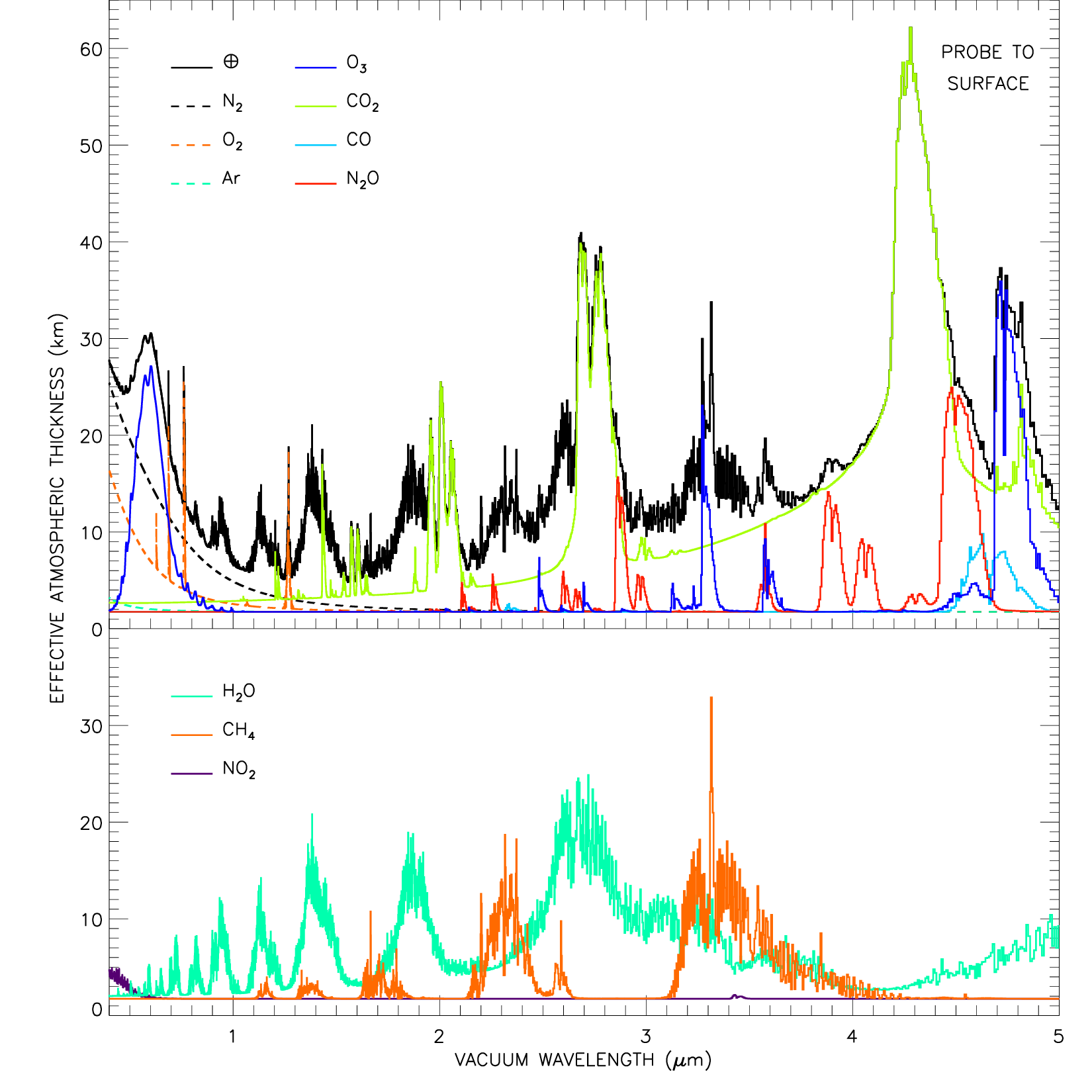}
\caption{Transmission spectrum including the effects of refraction (solid black line) of a cloud-free 
transiting exo-Earth orbiting an M5-M9 star, where the entire atmosphere can be probed ($\Delta z_{c} = 0$~km). 
Individual contribution from each of the important chemical species are also shown (see legend in the upper left corner of 
each panel).
\label{fig6}}
\end{figure}

Figure~\ref{fig6} shows the 0.4-5.0~$\mu$m transmission spectrum for an exo-Earth where the entire atmosphere can be 
probed, hence orbiting M5-M9 stars. This simulation is for a cloud-free atmosphere with a STP 
refractivity of 2.88$\times10^{-4}$. The individual contribution 
of each species is also shown, computed by assuming that the species in question is the only one with a non-zero
opacity. Note that the impact parameter of the surface lies 1.74~km above the projected surface because of the mapping 
of $r_{min}$ into $b$, and therefore this is the minimum value that the effective atmospheric thickness can have 
even in the absence of any absorption. 
The strongest bands in the spectrum are due to CO$_2$ absorption and are centered around 2.75 and 4.30~$\mu$m. 
A triple-band CO$_2$ signature (1.96-2.06~$\mu$m), as well as a few other CO$_2$ bands (1.20, 1.22, 1.43, 1.53, 1.57, and 
1.60~$\mu$m), can be seen through the H$_2$O continuum. One other narrow CO$_2$ band centered around 4.82~$\mu$m, enhances 
the long-wavelength side of the strongest O$_3$ band (4.69-4.84~$\mu$m) in the spectrum. O$_3$ has another strong absorption 
feature from 0.50-0.68~$\mu$m with a series of small bands until 1.00~$\mu$m, but the latters are masked by H$_2$O features. 
There are a few narrow O$_3$ bands that can be seen above the background: a strong one at 3.27~$\mu$m, two weak ones at 2.48 and 
3.13~$\mu$m, and one intermediate one blended with an N$_2$O band around 3.56~$\mu$m. O$_2$ has even fewer bands, 
which can be seen above the Rayleigh scattering and H$_2$O background, at 0.69, 0.76, and 1.27~$\mu$m. N$_2$O has several bands blended
on the long-wavelength side (2.12, 2.87, and 4.50~$\mu$m) of much stronger CO$_2$ bands. There are a few broad N$_2$O peaks 
in the 3.87-4.12~$\mu$m region, but the wing of the strongest CO$_2$ feature reduce their contrast significantly. All these
features can be seen above a background composed of Rayleigh scattering (shortward of 1.3~$\mu$m), CH$_4$ features (1.62-1.74, 2.14-2.46,
and 3.17-3.87~$\mu$m), and H$_2$O features (0.80-1.62, 1.74-1.94, 2.47-3.17, and 3.63-3.87~$\mu$m). Lastly, NO$_2$ and CO
create some very weak features below 0.48~$\mu$m, and 4.57-4.64~$\mu$m, respectively.

\begin{figure}
\epsscale{1.0}
\plotone{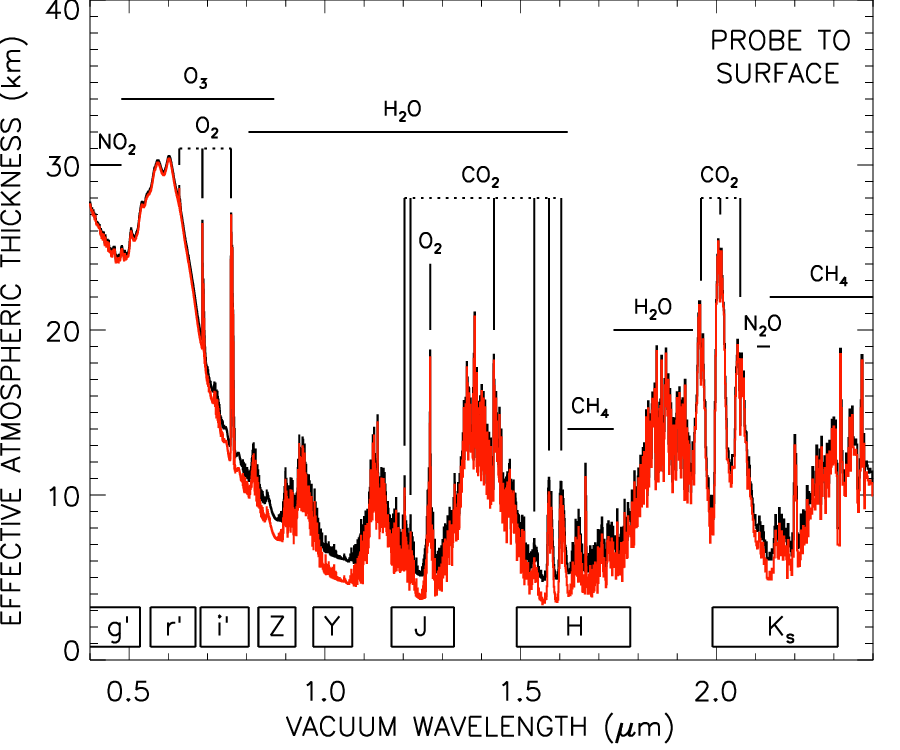}
\caption{VIS-NIR Transmission spectrum of a cloud-free transiting exo-Earth
orbiting an M5-M9 star ($\Delta z_{c} = 0$~km), with (black line) and without (red line) refraction. 
Spectral regions where the spectral signatures of a molecule are prominent are identified with solid horizontal lines.
Stronger individual bands of molecules are identified with solid vertical lines. Spectral coverage to the 50\% 
transmission limit of several commonly-used filters (see text) are indicated by the black rectangles. 
\label{fig7}}
\end{figure}

\begin{figure}
\epsscale{1.0}
\plotone{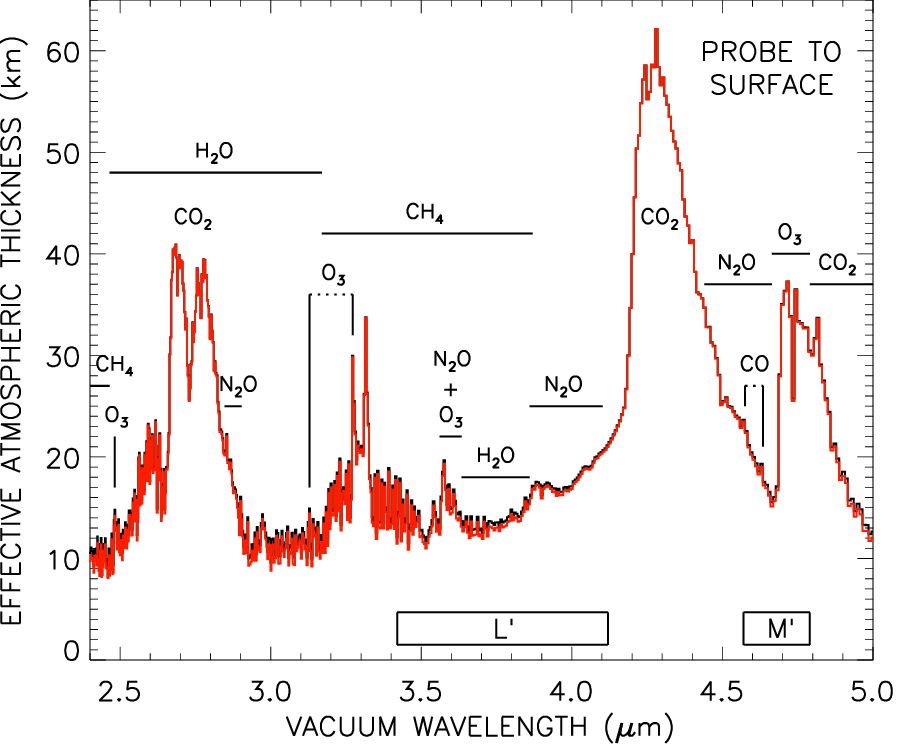}
\caption{IR Transmission spectrum of a cloud-free transiting exo-Earth orbiting an M5-M9 star ($\Delta z_{c} = 0$~km),
with (black line) and without (red line) refraction. See caption in Figure~\ref{fig7}.
\label{fig8}}
\end{figure}

\begin{figure}
\epsscale{1.0}
\plotone{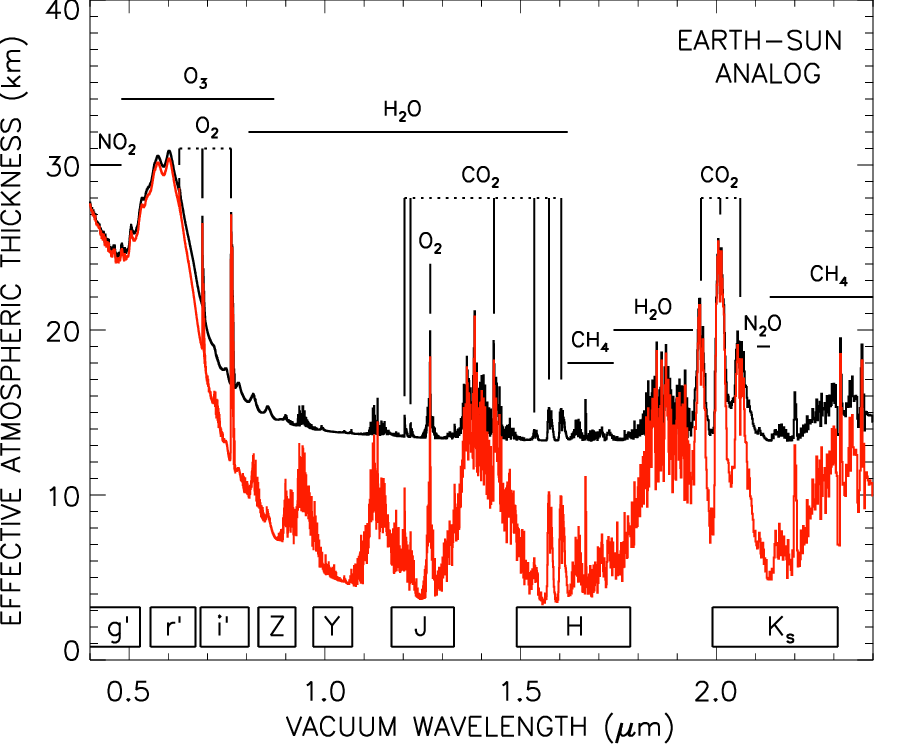}
\caption{VIS-NIR Transmission spectrum of a cloud-free transiting exo-Earth orbiting a Sun-like star 
($\Delta z_{c} = 12.7$~km), with (black line) and without (red line) refraction. 
See caption in Figure~\ref{fig7}.
\label{fig9}}
\end{figure}

\begin{figure}
\epsscale{1.0}
\plotone{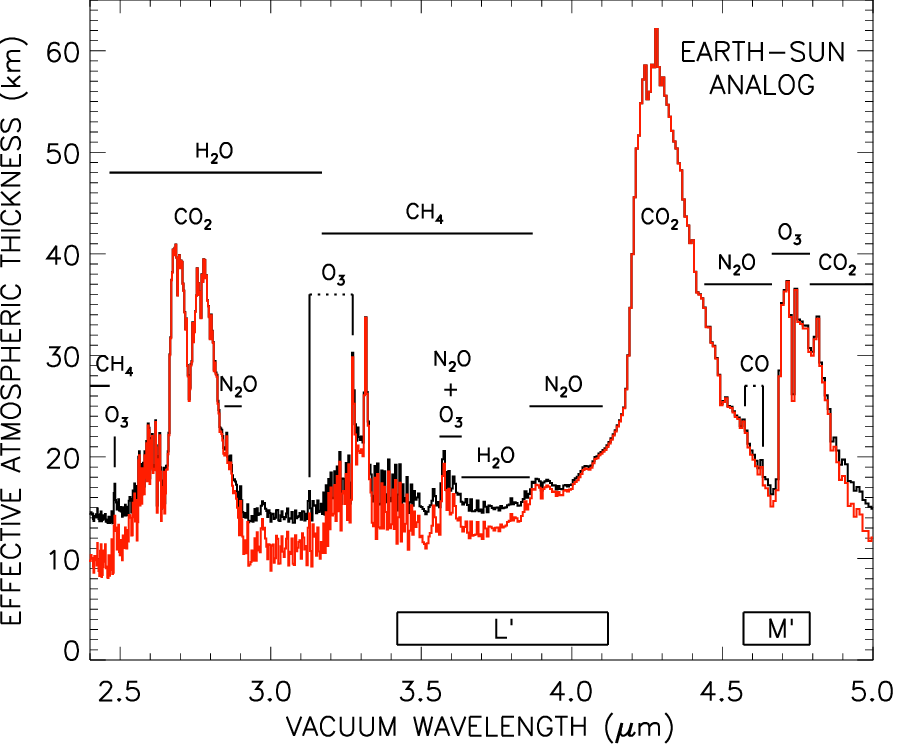}
\caption{IR Transmission spectrum of a cloud-free transiting exo-Earth orbiting a Sun-like star 
($\Delta z_{c} = 12.7$~km), with (black line) and without (red line) refraction.
See caption in Figure~\ref{fig7}.
\label{fig10}}
\end{figure}

Figures~\ref{fig7} and~\ref{fig8} show the impact of refraction on the transmission spectrum of an exo-Earth orbiting an M5-M9 star
for 0.4-2.4 and 2.4-5.0~$\mu$m, respectively. Figures~\ref{fig9} and~\ref{fig10} do the same but for an exo-Earth orbiting a 
Sun-like star. In all four figures, the refractive properties of the atmosphere are included or not by changing the STP
refractivity between 2.88$\times10^{-4}$ (black line) and 0 (red line). Note that for a refractiveless atmosphere ($\nu_{STP} = 0$), 
the critical altitude is always 0~km, irrespective of the star's spectral type. For reference, the spectral coverage of a 
few commonly-used filters in 
ground-based astronomy are also indicated to their 50\% transmission limit (see the review by Bessell 2005). These are
the g', r', and i' filters from the Sloan Digital Sky Survey (SDSS), the Z and Y filters from the UK Infrared Deep Sky Survey (UKIDSS), 
and the J, H, K$_s$, L', and M' filters from the Mauna Kea Observatories (MKO). 

Figures~\ref{fig7} to~\ref{fig10} show that, for an exo-Earth, most of the interesting spectral features fall outside of
the bandpass of commonly-used filters from the ground, as the Earth's atmosphere is not completely transparent at these 
wavelengths. The Y band probes a fairly featureless spectral region of an exo-Earth's transmission spectrum against which 
the other bands can be compared. The g', i', and Z bands sample the transmission spectrum where Rayleigh scattering 
is important (with some small contamination from other species), and the r' band can be used to look for O$_3$, even in 
photometric mode. The M' band also catches the strongest O$_3$ peak, with some minor contribution from CO$_2$ and CO. 
No unambiguous information can be obtained from the other bands in photometric mode, as more than one molecule 
falls inside their bandpass. If spectroscopy is possible, one can look for O$_2$ features in both the i' and the J bands.
The J band also includes CO$_2$, and H$_2$O signatures, while the H and K$_s$ bands covers 
mostly CO$_2$ and CH$_4$ signatures with a little contamination from H$_2$O and N$_2$O, respectively. The L' band samples the 
greatest variety of molecules: CO$_2$, CH$_4$, H$_2$O, N$_2$O, and O$_3$.

When an exo-Earth orbits an M5 to an M9 star, and the entire atmosphere can be probed (Figures~\ref{fig7} and~\ref{fig8}), 
refraction only increases the background of the spectrum at most by about 1.5~km, and the contrast in the 
molecular signatures are only slightly decreased. Most of the impact of refraction comes from the mapping 
of $r_{min}$ into $b$, rather than the increase in optical depth from the ray curvature. Indeed, the difference between 
$b$ and $r_{min}$ is 1.74, 0.94, and 0.44~km for rays reaching a minimum altitude of 0.0, 6.0, and 12.0~km, respectively. 
Conversely, the difference in atmospheric column abundances for these same rays is only about 7, 4, and 3\%, respectively. 
Considering the atmospheric scale height, this corresponds to an altitude change of 0.62, 0.36, and 0.12~km, smaller than the 
respective difference between $b$ and $r_{min}$. 

The geometric effect of refraction can have a much larger impact on the transmission spectrum than the other two 
effects. One must compare the effective atmospheric thickness of the transmission spectrum in the case where all of 
the atmosphere is probed, against the critical altitude associated with the planet-star system. In the spectral regions 
where the effective atmospheric thickness is significantly larger than the critical altitude, refraction will have only a 
minimal to no impact on the transmission spectrum because the lower atmospheric regions are dark irrespective of refraction. 
Conversely, spectral regions with effective atmospheric thickness below the critical altitude will see their effective 
atmospheric thickness increases to the critical altitude or above, with a decrease in the contrast of spectral features
as the observable column abundance of the atmosphere decreases. 

In the case of an Earth-Sun analog 
(Figures~\ref{fig9} and~\ref{fig10}) where only altitudes above 12.7~km are probed, many of the spectral features are 
masked or substantially reduced. Water absorption bands are severely affected as most of the H$_2$O is 
located below that altitude.
Conversely, the majority of O$_3$ is above that altitude. Hence, shortward of about 0.89~$\mu$m, all H$_2$O features are masked
by O$_3$ features. Below 1.2~$\mu$m, only the top of the H$_2$O bands around 0.94 and 1.13~$\mu$m can still be 
seen. Between 1.20 and 3.75~$\mu$m, many of the spectral features have their contrast reduced by a factor of 2 to 3, as the 
effective atmospheric thickness increases to a minimum background value of 13.2~km. The CO$_2$ band around 2.75~$\mu$m, as well 
as the 3.75-5.0~$\mu$m region, are mostly unaffected, as the atmosphere is opaque below the critical 
altitude. 

\begin{figure}
\epsscale{1.0}
\plotone{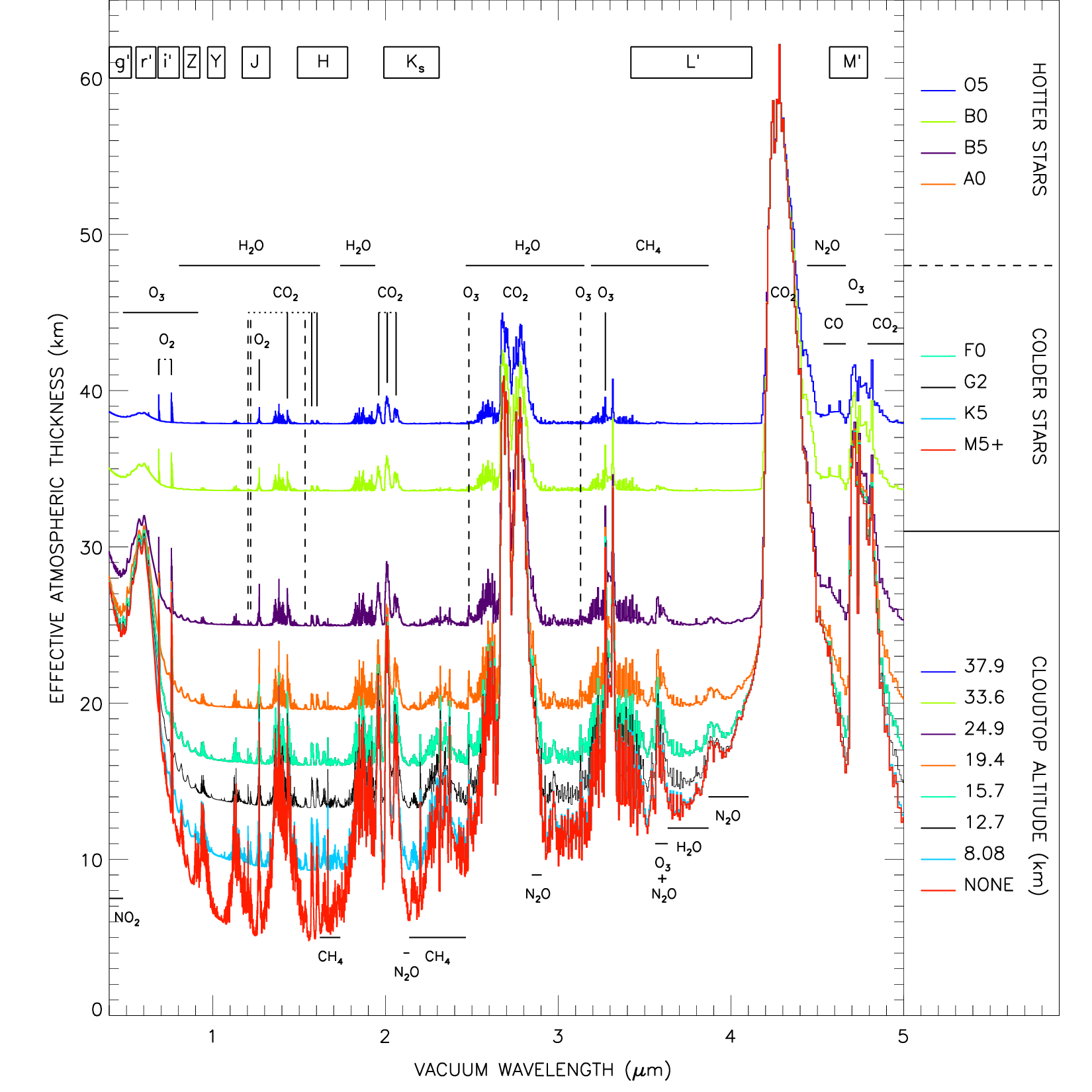}
\caption{VIS-IR Transmission spectra of a cloud-free transiting exo-Earth, that 
receives the same wavelength-integrated stellar flux as the Earth, orbiting stars of different 
spectral types (top right legend). A transiting exo-Earth orbiting an M5-M9 star,
with optically thick cloud up to the listed cloud-top altitudes
(bottom right legend), produces identical transmission spectra. 
Regions where the spectral signatures of a molecule are prominent are identified with solid 
horizontal lines. Stronger individual bands of other molecules are identified with solid vertical lines,  
except where they can not be observed in all stellar spectral type. In the latter case, a dashed vertical lines 
is drawn to the hottest star where they can be observed. Commonly-used filters are also 
indicated with black rectangles.
\label{fig11}}
\end{figure}

Around hotter stars, these effects become even more pronounced. Figure~\ref{fig11} shows transmission spectra of 
exo-Earths, that receive the same wavelength-integrated flux as the Earth, orbiting different spectral type stars. 
As the host star becomes hotter, the background of the spectrum increases, and the contrast of various 
spectral features decreases until for O5 stars, the hottest stars that we have considered, the exo-Earth transmission 
spectrum is flat with only a few remaining spectral features. Since CO$_2$ and O$_2$ are the most vertically mixed 
molecules with identifiable spectral features, most of their features can still be seen, albeit with a much reduced contrast. 
The only exceptions are the weaker CO$_2$ features at 1.20, 1.22, and 1.53~$\mu$m which become undetectable
for stars hotter than B0. The strongest CO$_2$ bands centered at 2.75 and 4.30~$\mu$m remain quite prominent for all 
stars. Although all O$_2$ features (0.69, 0.76, and 1.27~$\mu$m) are reduced in contrast around hotter stars, their 
importance relative to spectral features of other species actually increases (except for CO$_2$). Indeed, the 
nearby strong O$_3$ band centered at 0.59~$\mu$m, which is stronger than the O$_2$ features for stars as hot as B5 stars, 
is hardly detectable for an O5 star. The weak O$_3$ bands below 1.00~$\mu$m, as well as the weak bands at 2.48 and 3.13~$\mu$m
disappear for a B0 star while the 3.56~$\mu$m O$_3$ band is hardly detectable around an O5 star. The stronger 
bands at 3.27 and around 4.76~$\mu$m are observable for all stars. The relative contribution of the CO bands around 
4.60~$\mu$m increases with the temperature of the host star because CO is more abundant at altitudes above 50~km. 
N$_2$O features, which are weak throughout most of the spectrum, become undetectable around B0 host stars, except 
for the strongest band at 4.50~$\mu$m. Weaker CH$_4$ bands (1.62-1.74, 2.14-2.46, and 3.55-3.87~$\mu$m) are much reduced 
in contrast around an A0 star and become undetectable around B0 stars, while the stronger CH$_4$ band (3.17-3.55~$\mu$m) can be 
seen for all stars. H$_2$O band's strengths are reduced substantially from M5 to G2 stars, and only 3 bands (1.35-1.41, 1.80-1.95, 
and 3.20-3.64~$\mu$m) are still detectable around an O5 star. Finally, the slope of the spectrum shortward of 1.00~$\mu$m, 
where Rayleigh scattering is important, changes significantly with the spectral type of the star.

The change of the transmission spectrum with stellar spectral type is due to the change of
the lowest altitude probed, or the largest density and pressure probed. Although we have explored how the 
lowest altitude varies from the geometrical effect of refraction, other limiting factors exists. The presence of optically 
thick clouds, in this limb-viewing geometry, can blanket the lower atmospheric regions, and limit the lowest altitude, 
or highest density and pressure, probed to that of the cloud top. Hence, Figure~\ref{fig11} also shows the 
change in the transmission spectrum with the topmost altitude of a full cover of opaque clouds (see legend). 
The largest density probed can also be limited by the gas content of the atmosphere, whereby a more 
tenuous atmosphere will have a lower density at the surface. The largest density probed in a planet's atmosphere is determined from the 
lowest value from these three causes. If the clouds are not optically thick and let some light through, even in 
this limb-viewing geometry, then the transmission spectrum might have some exploitable signature of cloud optical 
properties. If refraction is the limiting factor, lower atmospheric regions should leave 
some signature in the light-curve profile, provided that they are not opaque. 

In Section~\ref{sedsect}, we discussed how potential temperature differences between the profile of 
an Earth-like planet and the Earth's atmospheric model, used in this paper, might impact the critical altitude. 
Since the transmission spectrum, expressed in terms of its effective atmospheric thickness, scales with 
scale height just as the critical altitude does to first order (see Section~\ref{sedsect}), an increase in 
scale height due to temperature changes will increase the scale rather than the shape 
of the spectral features. The second-order effect (see Equ.~\ref{critdens}) 
where higher densities can be probed with higher temperatures, imply that the background of the spectrum
will decrease very slightly as the temperature increases, and this effect is much smaller than the first. 
Hence, as temperatures increase, we expect the spectrum to scale with planetary temperatures, 
accompanied by a very slight increase in the contrast of the spectral features.

\section{Conclusions}

Atmospheric refraction plays a major role in transmission spectroscopy of habitable exoplanets. 
The geometric effect of refraction limits how deeply observers can probe a planet's atmospheres. The angular size 
of the host star with respect to the exoplanet, and to a lesser extent the size of the exoplanet itself, determine the 
critical deflection by which a ray can be bent and still reach the observer. Comparison of this critical deflection against 
the vertical bending properties of the atmosphere defines a critical altitude below which,
and a critical density and pressure above which, the atmosphere cannot be probed. 
When the critical deflection is larger than the deflection at the surface of the planet, observers can probe the atmosphere 
down to the surface. To first order, the critical deflection scales with $1/T_{\ast}^2$ (effective temperature of 
the host star) for exoplanets with identical incident stellar flux on the top of their atmospheres. Thus, observers can 
probe deeper in the atmosphere of planets orbiting cooler stars, for a given atmospheric profile.

The critical altitude and density depend on the composition of the atmosphere. CO$_2$ bends light more than 
an Earth-like N$_2$/O$_2$ atmosphere, and a Jupiter-like H$_2$/He atmosphere bends light the least. The large 
radius of gaseous giants can mitigate that effect. Also, super-Earths cannot be probed to densities as high as 
on Earth-sized planets or on mini-Neptunes.

We determined the critical altitudes, densities, and pressures for exoplanets with
identical mass, radius, and 1 bar N$_2$/O$_2$ atmosphere as the Earth, which receive the same 
incident stellar fluxes as Mars, Earth, and Venus, orbiting various stars from O5 to M9 spectral type. 
The hottest star around which all of this exo-Earth's atmosphere can be probed is an M1, M4, and M7, for 
the same incident stellar flux as Venus, Earth, and Mars, respectively. Transmission variability cannot be caused 
by clouds if the critical altitude is greater than the altitude of the topmost clouds. For an exo-Earth with an 
Earth-like incident stellar flux, this occurs around stars like the Sun or hotter. 

We illustrated the impact of refraction on the 0.4-5.0 $\mu$m transmission spectrum of an exo-Earth, receiving the same 
incident stellar flux as Earth, orbiting O5 to M9 stars. As the critical altitude increases, the spectrum's background
increases, decreasing the contrast of spectral features or masking them entirely. If an exo-Earth can exist 
around an O5 star, its transmission spectrum would be sparsely populated with only the strongest 
CO$_2$, H$_2$O, O$_2$, O$_3$, CH$_4$, CO, and N$_2$O bands, but would be otherwise flat, in stark contrast 
to that around M5 to M9 stars. The strongest bands are the 2.75 and 4.30~$\mu$m CO$_2$ bands and the 4.76~$\mu$m O$_3$ band. 
Whether the lowest altitude that can be probed is due to the refraction effect, or to the presence of optically thick
clouds in a limb-viewing geometry extending to the same altitude, or to a surface at the corresponding pressure,
makes no difference to the transmission spectrum. Hence, the geometrical effect of refraction is another phenomenon that 
can potentially explain flat transmission spectra.



\acknowledgments{The authors acknowledge support from DFG funding ENP Ka 3142/1-1 and 
the Simons Foundation on the Origins of Life Initiative.}

\clearpage



\clearpage

\clearpage

\clearpage

\clearpage

\clearpage

\clearpage

\clearpage

\clearpage

\clearpage

\clearpage

\clearpage







\clearpage 

\begin{deluxetable}{ccc}
\tabletypesize{\scriptsize}
\tablecaption{Effective Temperature and Radius of Main Sequence Stars. \label{tbl_star}}
\tablewidth{0pt}
\tablehead{
\colhead{ Stellar Spectral Type } & \colhead{ T$_{\ast}$ (K) } 
& \colhead{ R$_{\ast}$/R$_{\Sun}$}
}
\startdata
O5 & 42000 & 12.0 \\
B0 & 30000 & 7.4 \\
B5 & 15200 & 3.9 \\
B8 & 11400 & 3.0 \\
A0 & 9790 & 2.4 \\
A5 & 8180 & 1.7 \\
F0 & 7300 & 1.5 \\
F5 & 6650 & 1.3 \\
G0 & 5940 & 1.1 \\
$\Sun$ - G2 & 5778\tablenotemark{a} & 1.0 \\
G5 & 5560 & 0.92 \\
K0 & 5150 & 0.85 \\
K5 & 4410 & 0.72 \\
M0 & 3800 & 0.62 \\
M2 & 3400 & 0.44 \\
M4 & 3100 & 0.26\tablenotemark{b} \\
M5 & 2800 & 0.20 \\
M7 & 2500 & 0.12 \\
M9 & 2300 & 0.08 \\
\enddata
\tablecomments{Source: Reid \& Hawley (2005) for M~stars, Cox (2000) otherwise}
\tablenotetext{a}{{L}odders \& Fegley (1998)}
\tablenotetext{b}{Kaltenegger \& Traub (2009)}
\end{deluxetable}

\clearpage 

\begin{deluxetable}{cccccccccc}
\tabletypesize{\scriptsize}
\tablecaption{Exo-Earth Critical Altitude, Density, and Pressure vs. Stellar Spectral Type \label{tbl_critalt}}
\tablecolumns{10}
\tablewidth{0pt}
\tablehead{ 
\multicolumn{1}{c}{Stellar} & \multicolumn{3}{c}{Critical Altitude (km)} & \multicolumn{3}{c}{Critical Density (amagat)}
& \multicolumn{3}{c}{Critical Pressure (mbar)} \\
\multicolumn{1}{c}{Spectral} & \multicolumn{3}{c}{for same incident flux as} & \multicolumn{3}{c}{for same incident flux as}
& \multicolumn{3}{c}{for same incident flux as} \\
\multicolumn{1}{c}{Type} & \multicolumn{1}{c}{Venus} & \multicolumn{1}{c}{Earth} & \multicolumn{1}{c}{Mars} &
\multicolumn{1}{c}{Venus} & \multicolumn{1}{c}{Earth} & \multicolumn{1}{c}{Mars} &
\multicolumn{1}{c}{Venus} & \multicolumn{1}{c}{Earth} & \multicolumn{1}{c}{Mars}
}
\startdata
O5 & 35.7 & 37.9 & 40.6 & 5.84$\times$10$^{-3}$ & 4.23$\times$10$^{-3}$ & 2.84$\times$10$^{-3}$
                        & 5.17 & 3.84 & 2.65 \\
B0 & 31.5 & 33.6 & 36.2 & 1.13$\times$10$^{-2}$ & 8.19$\times$10$^{-3}$ & 5.46$\times$10$^{-3}$
                        & 9.54 & 7.06 & 4.85 \\
B5 & 22.9 & 24.9 & 27.5 & 4.34$\times$10$^{-2}$ & 3.16$\times$10$^{-2}$ & 2.09$\times$10$^{-2}$
                        & 35.3 & 26.0 & 17.3 \\
B8 & 19.3 & 21.3 & 23.9 & 7.70$\times$10$^{-2}$ & 5.56$\times$10$^{-2}$ & 3.67$\times$10$^{-2}$
                        & 61.9 & 45.0 & 30.1 \\
A0 & 17.3 & 19.4 & 22.0 & 1.05$\times$10$^{-1}$ & 7.56$\times$10$^{-2}$ & 4.96$\times$10$^{-2}$
                        & 84.0 & 60.7 & 40.2 \\
A5 & 15.1 & 17.1 & 19.8 & 1.49$\times$10$^{-1}$ & 1.08$\times$10$^{-1}$ & 7.10$\times$10$^{-2}$
                        & 120 & 87.1 & 57.1 \\
F0 & 13.6 & 15.7 & 18.3 & 1.86$\times$10$^{-1}$ & 1.36$\times$10$^{-1}$ & 8.95$\times$10$^{-2}$
                        & 150 & 109 & 71.9 \\
F5 & 12.5 & 14.5 & 17.1 & 2.24$\times$10$^{-1}$ & 1.63$\times$10$^{-1}$ & 1.08$\times$10$^{-1}$
                        & 180 & 131 & 86.6 \\
G0 & 11.1 & 13.1 & 15.7 & 2.80$\times$10$^{-1}$ & 2.04$\times$10$^{-1}$ & 1.35$\times$10$^{-1}$
                        & 225 & 164 & 108 \\
$\Sun$ - G2 & 10.2 & 12.7 & 15.4 & 3.10$\times$10$^{-1}$ & 2.15$\times$10$^{-1}$ & 1.42$\times$10$^{-1}$
                        & 255 & 173 & 114 \\
G5 & 9.44 & 12.2 & 14.9 & 3.43$\times$10$^{-1}$ & 2.32$\times$10$^{-1}$ & 1.54$\times$10$^{-1}$
                        & 288 & 187 & 123 \\
K0 & 7.99 & 11.3 & 13.9 & 4.07$\times$10$^{-1}$ & 2.70$\times$10$^{-1}$ & 1.79$\times$10$^{-1}$
                        & 357 & 217 & 144 \\
K5 & 4.90 & 8.08 & 12.0 & 5.76$\times$10$^{-1}$ & 4.03$\times$10$^{-1}$ & 2.43$\times$10$^{-1}$
                        & 548 & 353 & 195 \\
M0 & 1.78 & 5.15 & 9.24 & 7.96$\times$10$^{-1}$ & 5.61$\times$10$^{-1}$ & 3.51$\times$10$^{-1}$
                        & 817 & 530 & 297 \\
M2 & 0.00 & 2.79 & 7.06 & 9.48$\times$10$^{-1}$ & 7.12$\times$10$^{-1}$ & 4.53$\times$10$^{-1}$
                        & 1013 & 720 & 407 \\
M4 & 0.00 & 0.66 & 5.09 & 9.48$\times$10$^{-1}$ & 8.89$\times$10$^{-1}$ & 5.64$\times$10$^{-1}$
                        & 1013 & 936 & 534 \\
M5 & 0.00 & 0.00 & 2.89 & 9.48$\times$10$^{-1}$ & 9.48$\times$10$^{-1}$ & 7.12$\times$10$^{-1}$
                        & 1013 & 1013 & 711 \\
M7 & 0.00 & 0.00 & 0.15 & 9.48$\times$10$^{-1}$ & 9.48$\times$10$^{-1}$ & 9.34$\times$10$^{-1}$
                        & 1013 & 1013 & 995 \\
M9 & 0.00 & 0.00 & 0.00 & 9.48$\times$10$^{-1}$ & 9.48$\times$10$^{-1}$ & 9.48$\times$10$^{-1}$
                        & 1013 & 1013 & 1013 \\
\enddata
\tablecomments{The critical altitudes, densities, and pressures are listed for an 
exo-Earth which is at a distance from its star such that the stellar radiation at 
the top of its atmosphere is identical to that received by present-day Venus, Earth and Mars, 
respectively. The critical altitudes were computed on a meter-scale vertical resolution 
for a STP refractivity of $2.88\times10^{-4}$, and rounded to 0.1 or 0.01~km 
depending on the magnitude of the critical altitude.}
\end{deluxetable}







\end{document}